\documentclass[iop]{emulateapj}

\usepackage[colorlinks,urlcolor=blue,citecolor=blue,linkcolor=blue]{hyperref}

\shorttitle{Cosmic-ray-induced filamentation instability}
\shortauthors{Caprioli \& Spitkovsky}

\begin{document}

\title{Cosmic-ray--induced filamentation instability in collisionless shocks}

\author{D. Caprioli and A. Spitkovsky}
\affil{Department of Astrophysical Sciences, Princeton University, 
    4 Ivy Ln., Princeton NJ 08544}
\email{caprioli@astro.princeton.edu}

\begin{abstract}
We used unprecedentedly large 2D and 3D hybrid (kinetic ions -- fluid electrons) simulations of non-relativistic collisionless strong shocks in order to investigate the effects of self-consistently accelerated ions on the overall shock dynamics. 
The current driven by suprathermal particles streaming ahead of the shock excites modes transverse to the background magnetic field. 
The Lorentz force induced by these self-amplified fields tends to excavate tubular, underdense, magnetic-field-depleted cavities that are advected with the fluid and perturb the shock surface, triggering downstream turbulent motions. 
These motions further amplify the magnetic field, up to factors of 50--100 in knot-like structures.
Once downstream, the cavities tend to be filled by hot plasma plumes that compress and stretch the magnetic fields in elongated filaments;
this effect is particularly evident if the shock propagates parallel to the background field.
Highly-magnetized knots and filaments may provide explanations for the rapid X-ray variability observed in RX J1713.7-3946 and for the regular pattern of X-ray bright stripes detected in Tycho's supernova remnant.
\end{abstract}


\section{Introduction}
Following the pioneering idea of \cite{Fermi49}, in the late '70s several authors realized that  collisionless shocks are prominent sites for the acceleration of particles \citep{krymskii77,bell78a,blandford-ostriker78}.
More recently, the detection of narrow X-ray rims in young supernova remnants (SNRs), whose origin has been explained as synchrotron emission of relativistic electrons radiating in magnetic fields of a few hundreds $\mu$G, has provided evidence that the level of magnetization at SNR shocks is much larger than in the interstellar medium
\citep[see e.g.,][]{vink-laming03,Bamba+05,P+06}.

This association between particle acceleration and magnetic field amplification has been welcomed by theorists for several reasons.
First, the super-Alfv\'enic streaming of accelerated particles is predicted to excite different plasma instabilities \citep[e.g.,][]{bell78a,bell04}, which can account for the inferred levels of magnetization \citep[at least as an extrapolation of the linear theory, see, e.g.,][]{jumpkin}.
Second, the interstellar turbulence cannot scatter particles effectively enough to achieve the highest energies measured in Galactic cosmic rays, while self-generated magnetic fields can \citep[e.g.][]{bac07}. 
Finally, amplified magnetic fields may represent a key ingredient for explaining the steep spectra inferred from recent $\gamma$-ray observations \citep{efficiency}.

\begin{figure*}\centering
\includegraphics[trim=40px 20px 20px 40px, clip=true, width=1.00\textwidth]{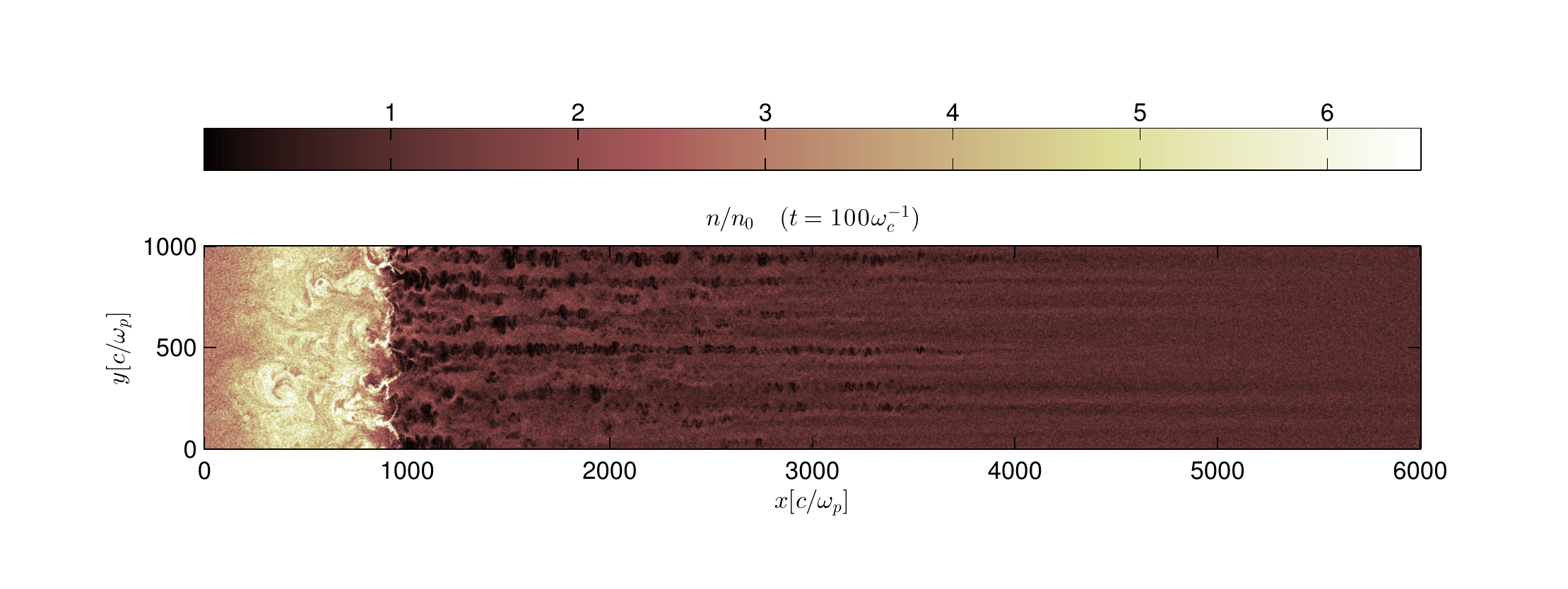}
\includegraphics[trim=40px 20px 20px 85px, clip=true, width=1.00\textwidth]{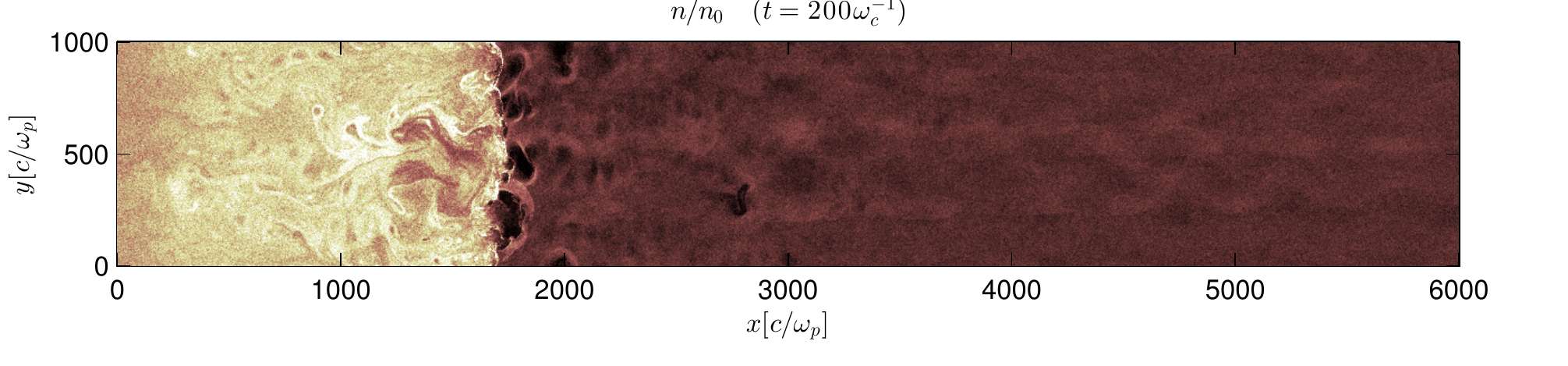}
\includegraphics[trim=40px 20px 20px 85px, clip=true, width=1.00\textwidth]{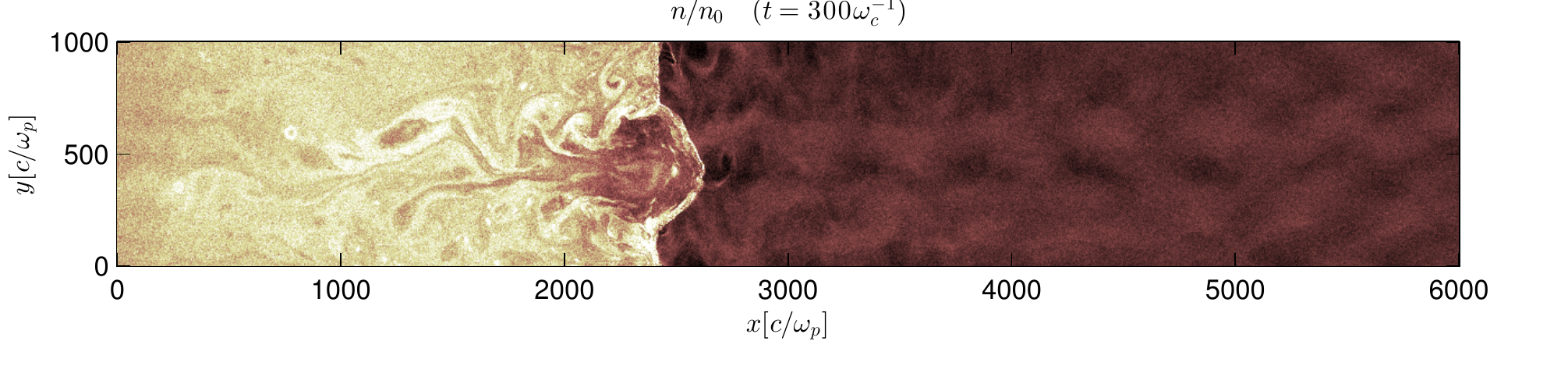}
\includegraphics[trim=40px 20px 20px 85px, clip=true, width=1.00\textwidth]{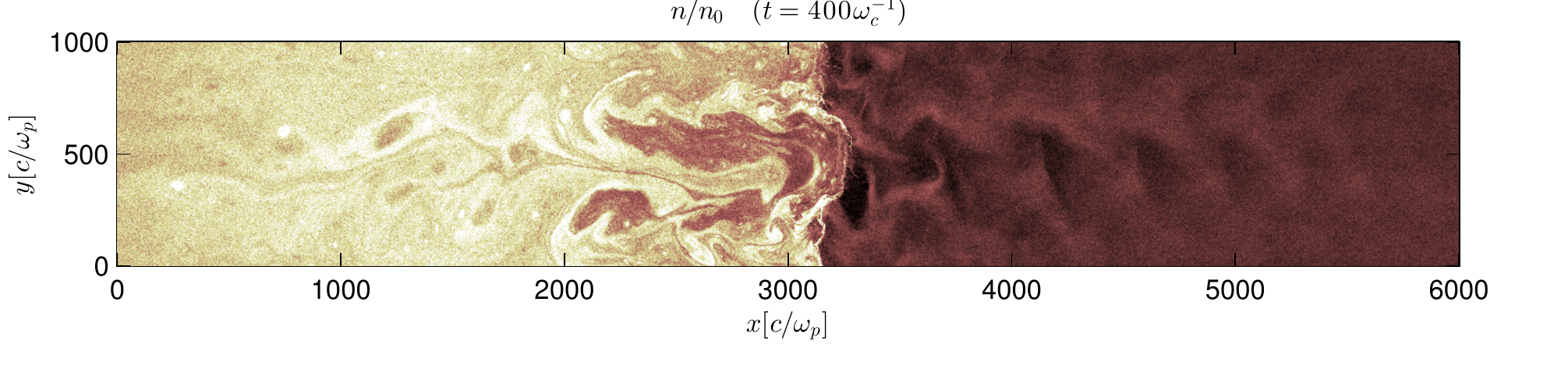}
\includegraphics[trim=40px 20px 20px 85px, clip=true, width=1.00\textwidth]{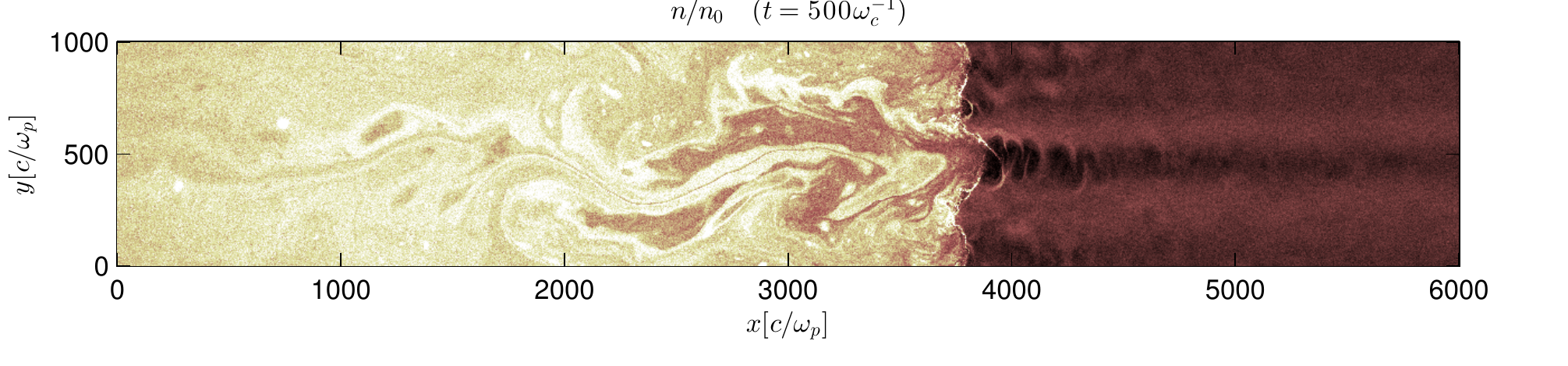}
\caption{\label{fig:rho}
Evolution of the plasma density for a 2D, parallel collisionless shock with Alfv\'enic Mach number $M_A=30$, as seen in the downstream frame. 
The total computational box measures $10^4\times10^3 (c/\omega_p)^2$, only a portion of which is showed here.
For comparison, the nominal gyroradius of an ion with velocity $v_{sh}$ is $r_L=v_{sh}/\omega_c=30c/\omega_p$ in the upstream magnetic field, while the most energetic particles at the end of the simulation have gyroradii as large as $r_L(E_{max})\sim 300c/\omega_p$.
\emph{An animation is available in the online journal}.
}
\end{figure*}

The details of such an interplay between accelerated particles and magnetic fields have not been completely understood yet, mostly because of the difficulty in accounting for the particle-wave coupling in the fully non-linear regime.
Nevertheless, collisionless shocks are mediated by electromagnetic interactions only, so they can be modeled by iteratively moving particles on a grid according to the Lorentz force and adjusting the electromagnetic configuration via Maxwell equations. 
Such a particle-in-cell (PIC) approach provides great insight into the properties of collisionless shocks \citep[see, e.g.,][]{stroman+09,ohira+09,rs09,rs10,ss11,Niemiec+12}.
However, PIC codes are computationally quite expensive, in that they require the plasma and gyration scales of both ions and electrons to be resolved. 
To partially mitigate this problem, one can adopt the so-called \emph{hybrid} approach, in which the ions, which drive the dynamics, are treated kinetically while the (massless) electrons are treated as a neutralizing fluid.
This approach neglects the small electron scales and allows the investigation of more macroscopic phenomena in much larger (in physical units) computational boxes.
In particular, it has been widely exploited to study ion acceleration at collisionless shocks in different astrophysical environments \citep[see, e.g.,][]{Winske85, Lipatov02, Giacalone+97,ge00,Giacalone04,gs12}. 

Here we show the results of 2D and 3D hybrid simulations run with the non-relativistic code \emph{dHybrid} \citep{gargate+07}.
These simulations allow for unprecedentedly large computational boxes, especially in the direction transverse to the shock velocity, probing the multi-dimensional structure of non-relativistic collisionless shocks.
We focus our attention on shocks with high-Mach numbers ($M\gtrsim 30$);
this regime is most relevant for SNR shocks, and so far has been scarcely studied because of the large dynamical range required to follow both thermal and non-thermal particles. 

We investigate, for the first time in a self-consistent simulation of a collisionless shock with accelerated particles, the formation  of peculiar structures in the upstream, which are shaped as tubular cavities surrounded by a net of dense filaments where the magnetic field is significantly amplified.
That accelerated particles drive an instability that is filamentary in nature has already been put forward by \cite{bell04,bell05}, who have run magneto-hydrodynamical simulations with a fixed current imposed through a periodic box seeded with an initial turbulence.
Recently, \cite{rb12} have provided an analytical derivation of the cavity growth rate, supporting their findings with simulations that couple a fluid treatment of the background plasma with a kinetic description of the energetic ions driving the instability. The instability was studied in a 2D slab geometry representing a section of the upstream perpendicular to the shock normal.

The simulations presented here differ from those in the literature in several respects:
1) the ion current is time-dependent and self-consistently generated by shock acceleration, and not estimated by assuming the relativistic ions to be isotropic in the shock frame, which translates into an anisotropy of order $\sim v_s/c$ in the upstream reference frame (here $v_s$ is the shock speed);
2) we do not need to seed the fluid with pre-existing turbulence, since accelerated particles generate it by themselves via streaming instability; 
3) in our setup filamentation develops while the fluid is advected towards the shock, so that we retain the evolution and the 3D features of cavities and filaments;
4) simulating the global shock structure allows us to show how, when advected through the shock, cavities and filaments affect the nature of the discontinuity and the magnetic field topology.

In \S\ref{sec:sims} we outline the main features of our 2D and 3D simulations for both parallel and oblique shocks, and discuss the mechanisms that lead to the formation of cavities and filaments. 
In \S\ref{sec:obs} we discuss some possible observational implications of such filamentation in the context of the synchrotron emission detected in young SNRs like Tycho and RX J1713.7-3946.

\section{Hybrid simulations}\label{sec:sims}

\begin{figure*}\centering
\includegraphics[trim=20px 50px 0px 0px, clip=true, width=.80\textwidth]{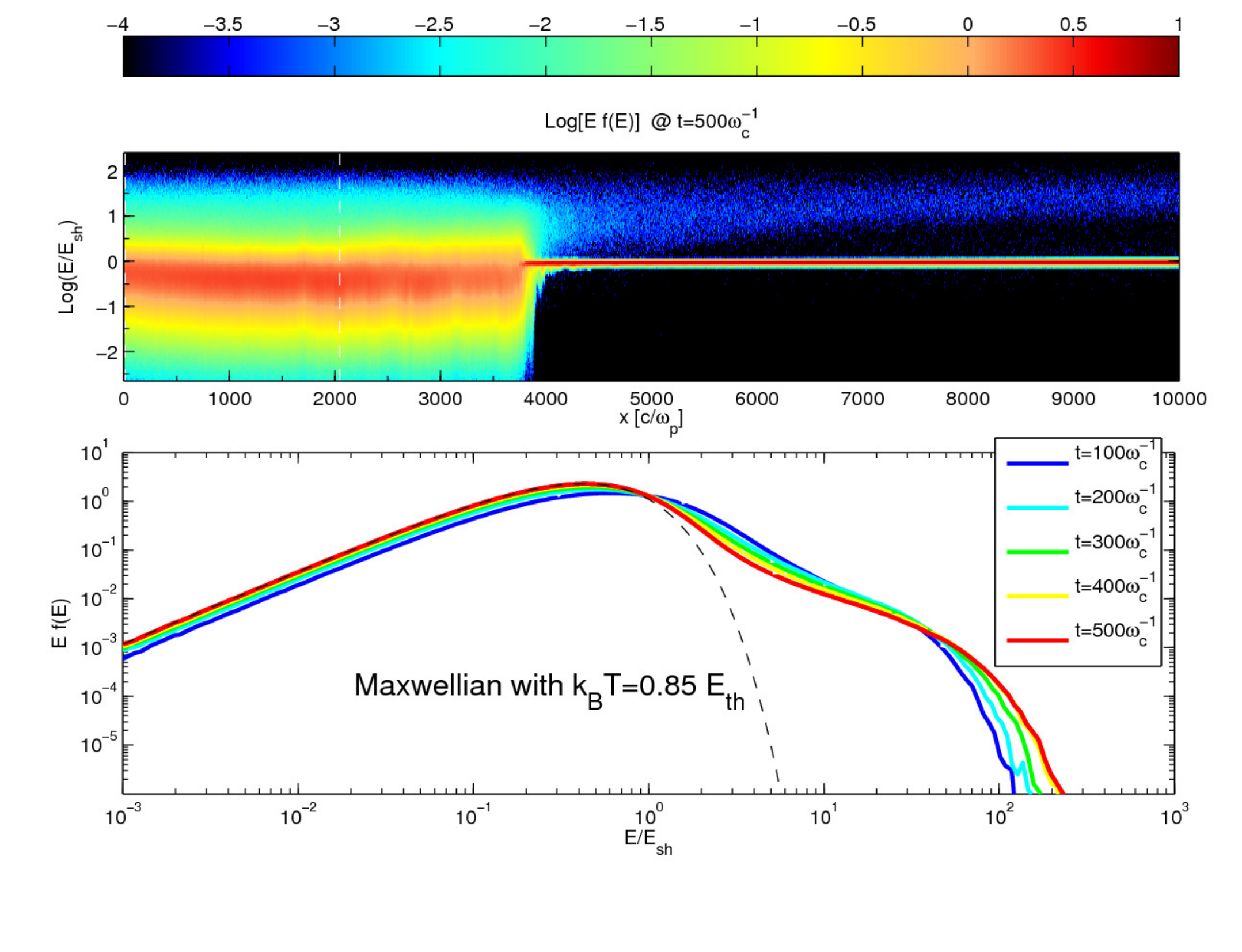}
\caption{\label{fig:spectra}
\emph{Top panel:} Particle energy spectrum $f(E)$ as a function of $x$ in units of $E_{sh}=mv_{sh}^2/2$ at time $t=500\omega_c^{-1}$.
\emph{Bottom panel:} Time evolution of the particle spectrum for $x<2000c/\omega_p$, as seen in the downstream reference frame.
For comparison, the Maxwellian distribution corresponding to a temperature $T=0.85 E_{th}/k_B$ is showed as well (dashed line), with $E_{th}=3/8E_{sh}$ the temperature given by standard jump conditions.} 
\end{figure*}

\begin{figure*}\centering
\includegraphics[trim=40px 50px 40px 55px, clip=true, width=.80\textwidth]{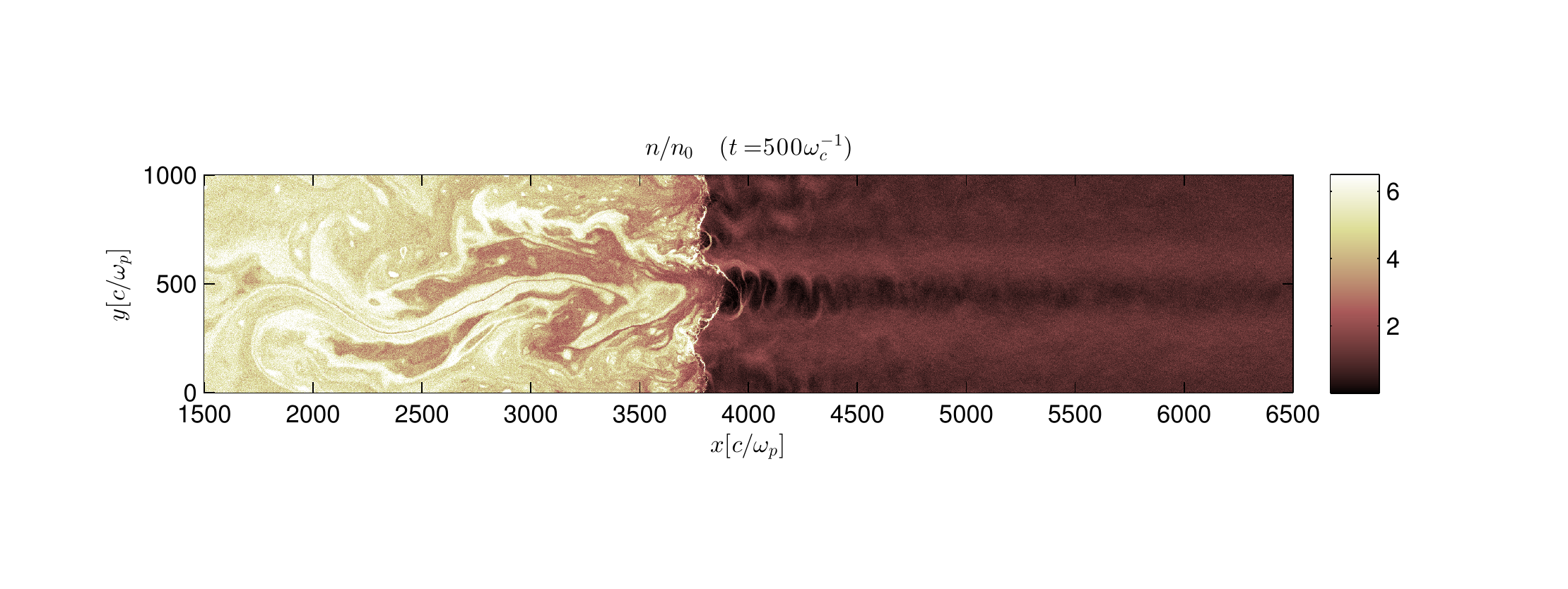}
\includegraphics[trim=40px 50px 40px 50px, clip=true, width=.80\textwidth]{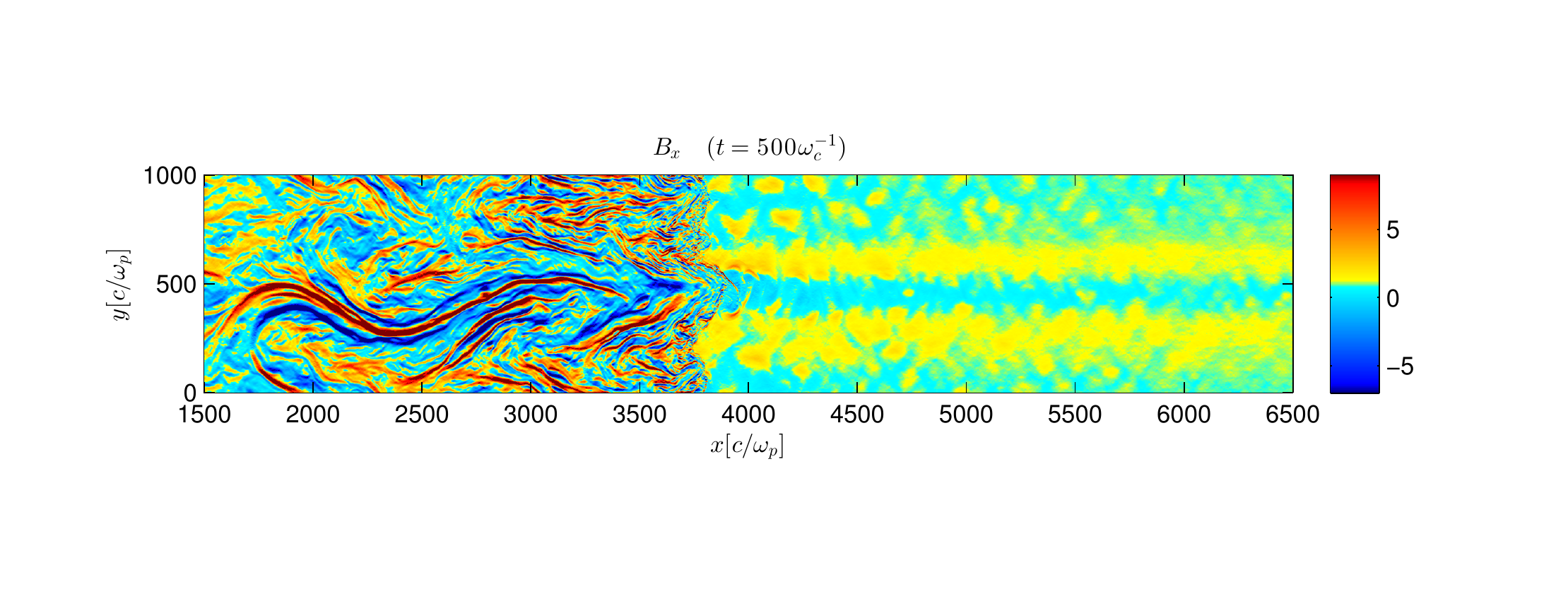}
\includegraphics[trim=40px 50px 40px 50px, clip=true, width=.80\textwidth]{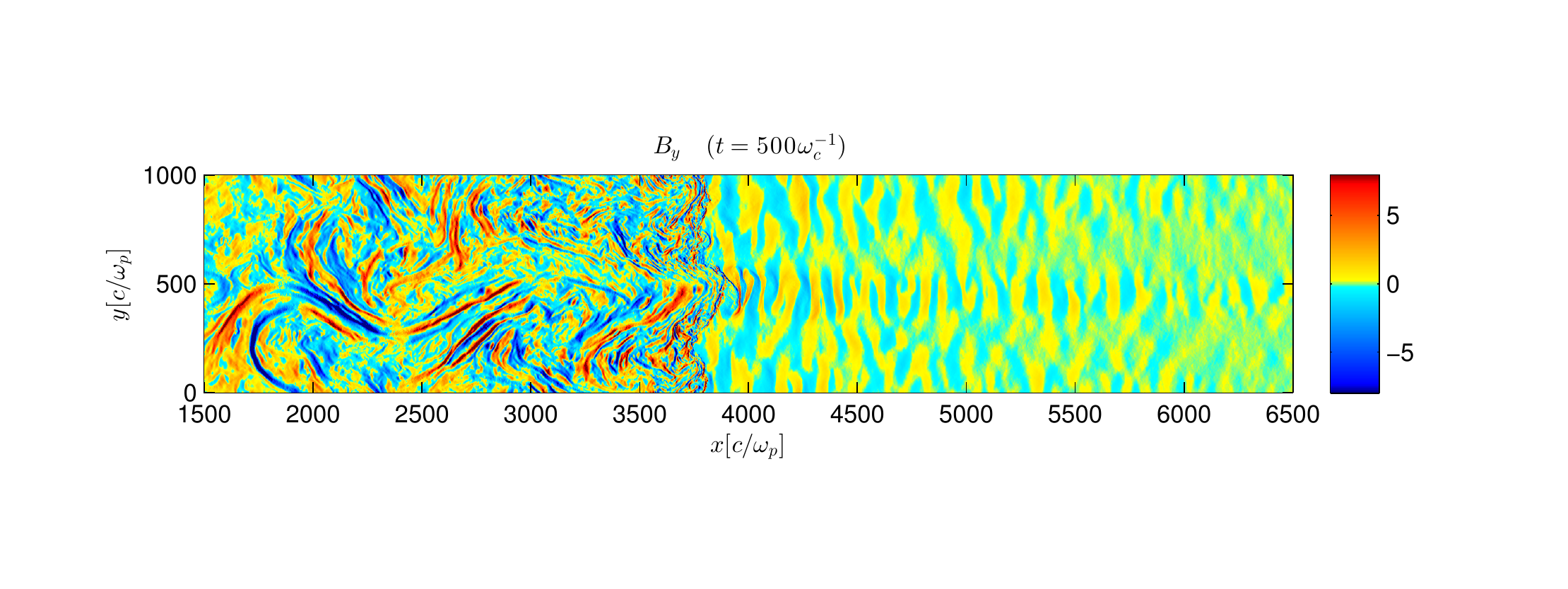}
\includegraphics[trim=40px 50px 40px 50px, clip=true, width=.80\textwidth]{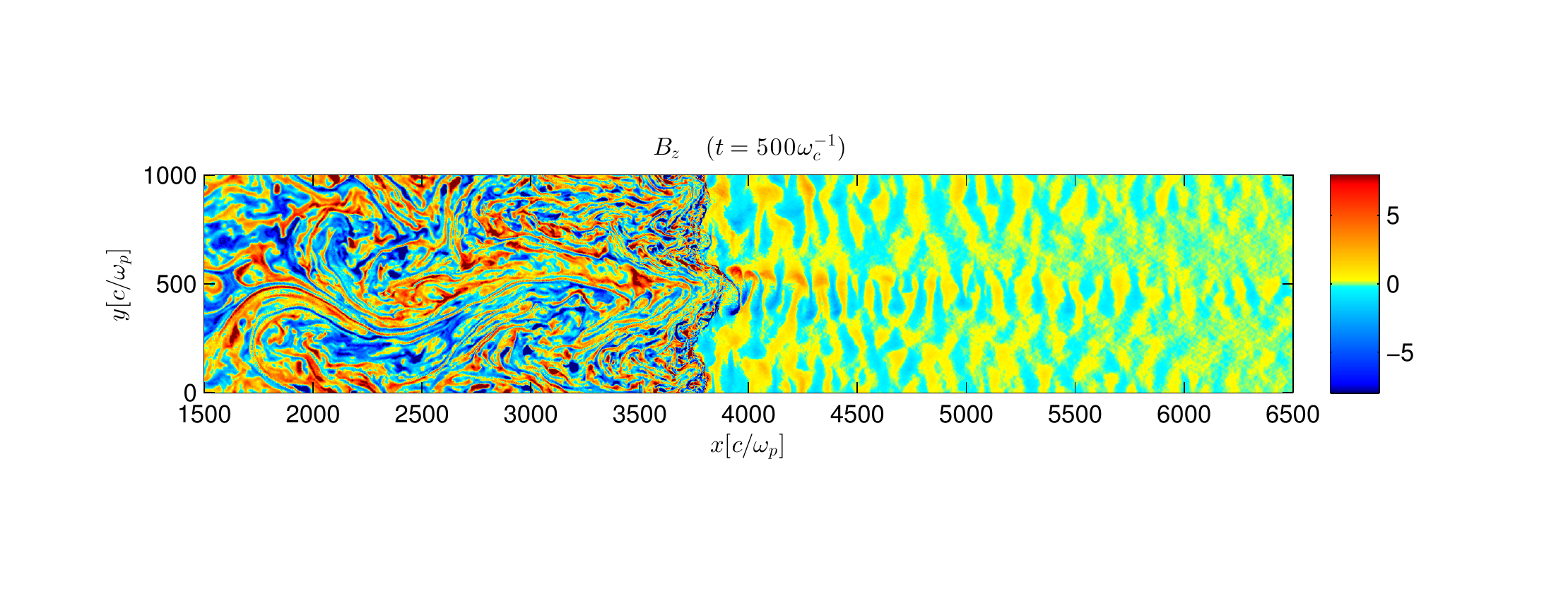}
\includegraphics[trim=40px 50px 40px 50px, clip=true, width=.80\textwidth]{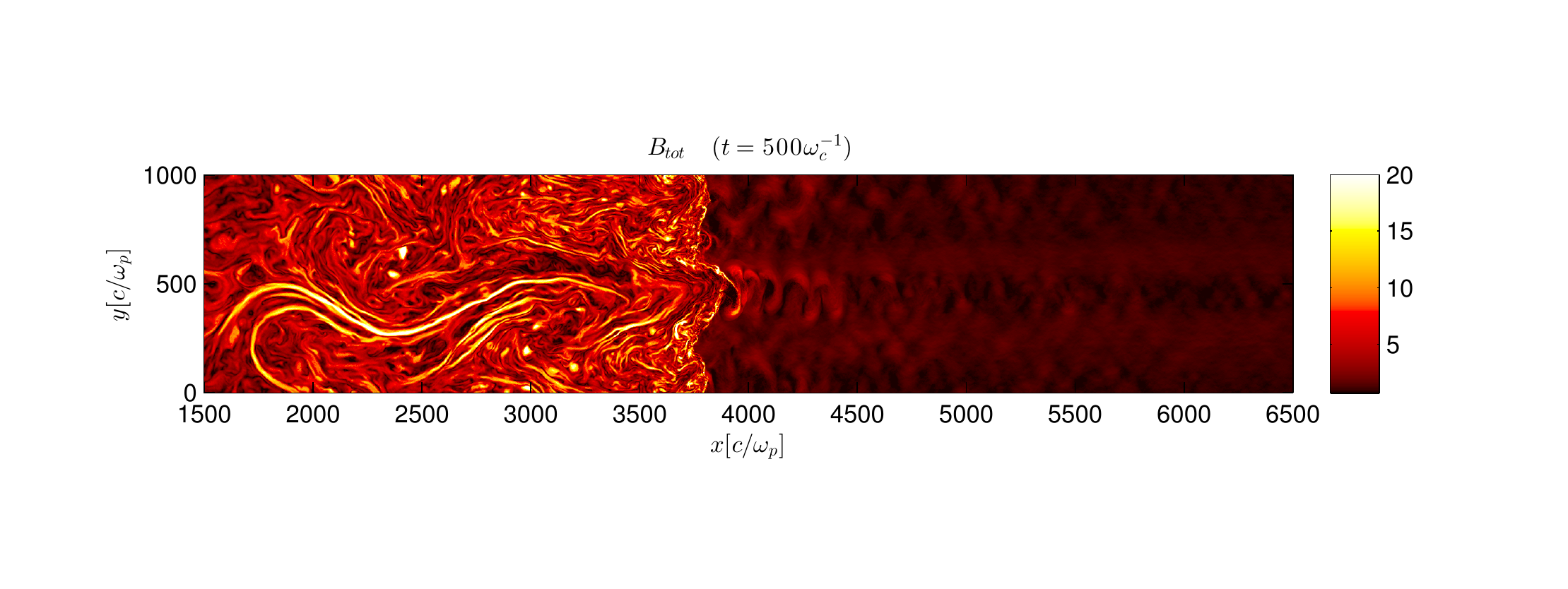}
\includegraphics[trim=40px 50px 40px 50px, clip=true, width=.80\textwidth]{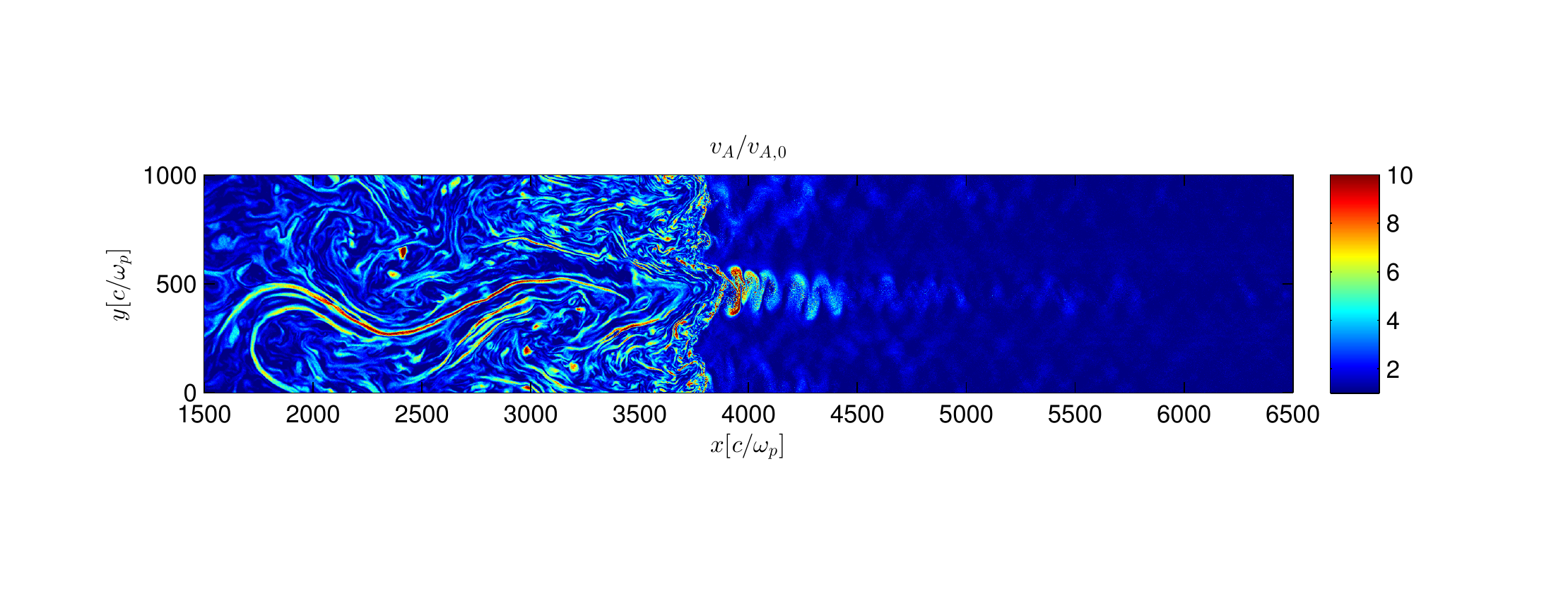}
\caption{\label{fig:par}
Density, parallel ($B_x$), transverse ($B_y$,$B_z$) and total ($B_{tot}$) magnetic field, and Alfv\'en velocity $v_A=B_{tot}/\sqrt{4\pi mn}$ for the 2D simulation in Figure~\ref{fig:rho} at $t=500\omega_c^{-1}$. 
All the quantities are normalized to their initial values.
A filament with $B_{tot}\approx 15-20B_0$ and some knots where $B_{tot}\approx 20-40B_0$ are clearly visible. \emph{An animation is available in the online journal}.}
\end{figure*}

\begin{figure*}\centering
\includegraphics[trim=40px 50px 40px 55px, clip=true, width=.80\textwidth]{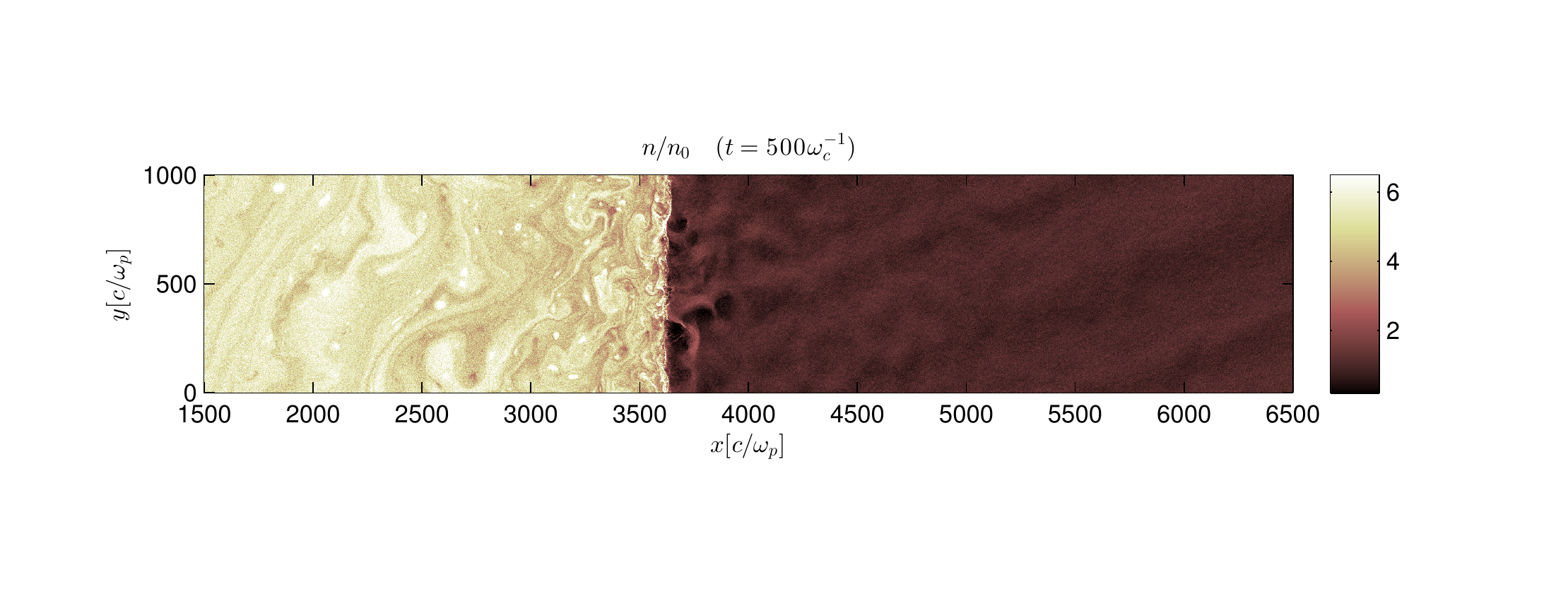}
\includegraphics[trim=40px 50px 40px 50px, clip=true, width=.80\textwidth]{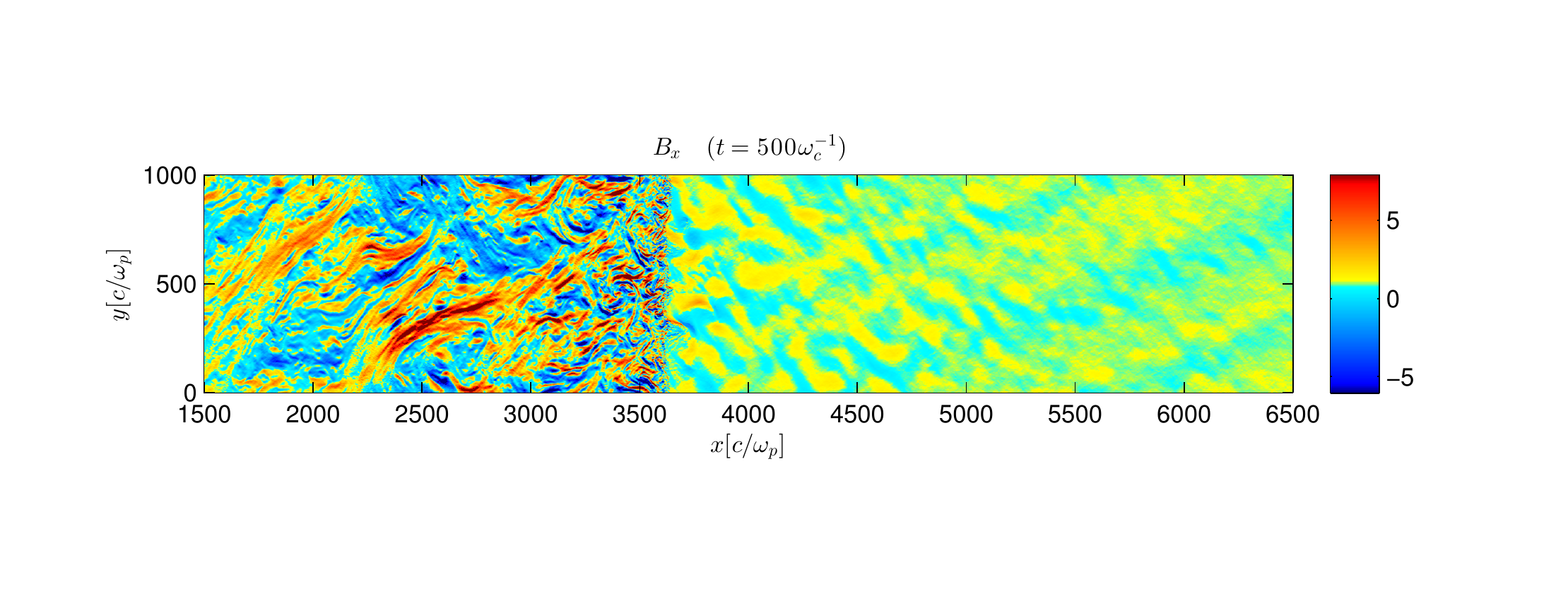}
\includegraphics[trim=40px 50px 40px 50px, clip=true, width=.80\textwidth]{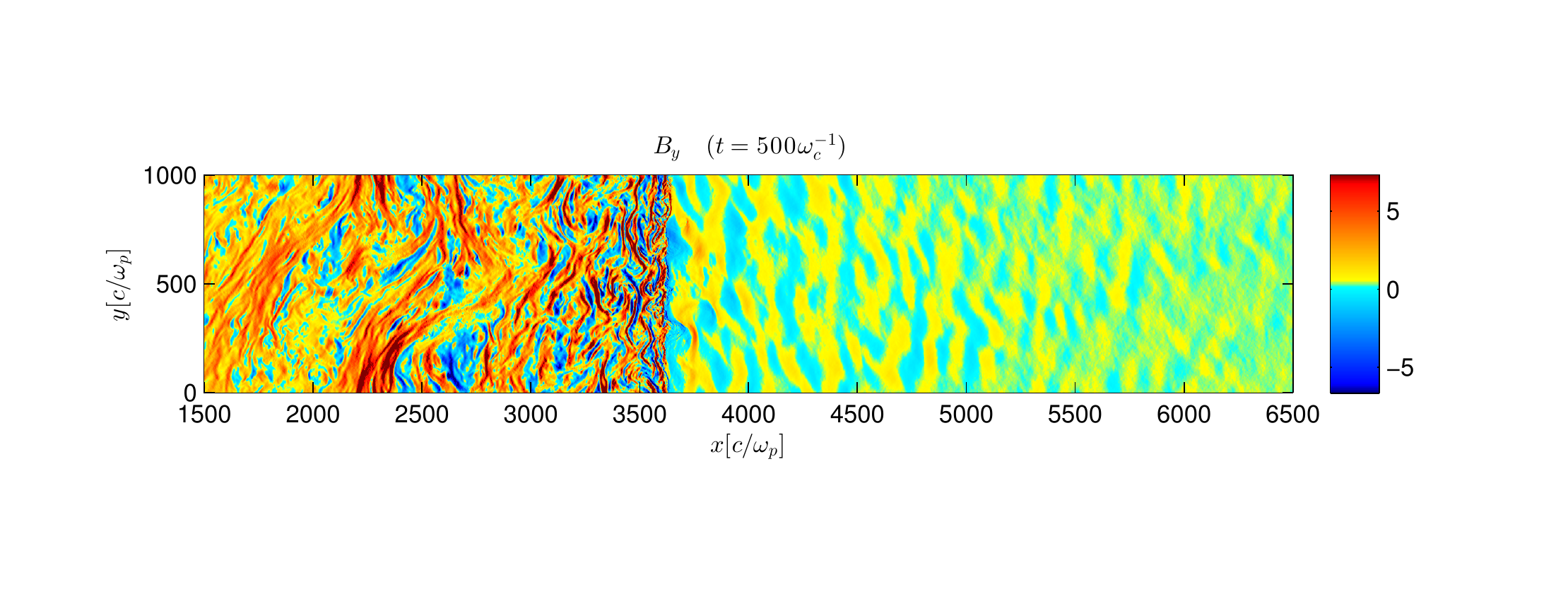}
\includegraphics[trim=40px 50px 40px 50px, clip=true, width=.80\textwidth]{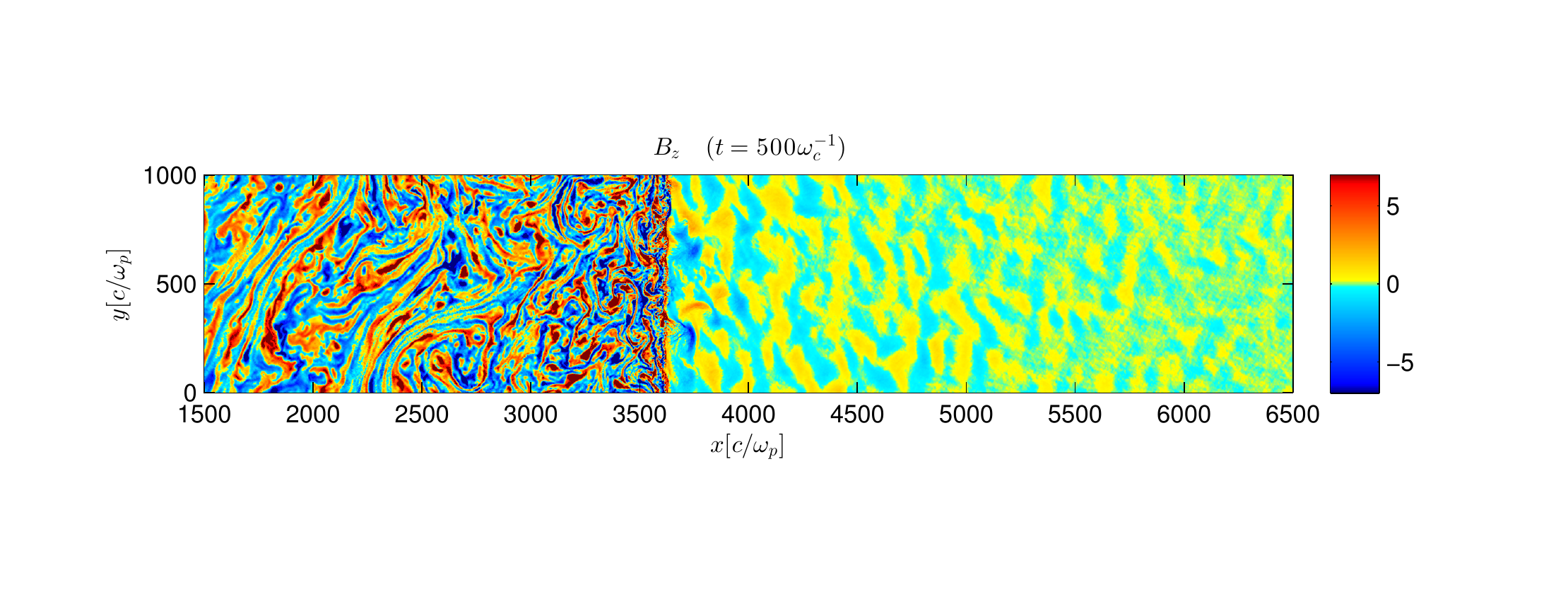}
\includegraphics[trim=40px 50px 40px 50px, clip=true, width=.80\textwidth]{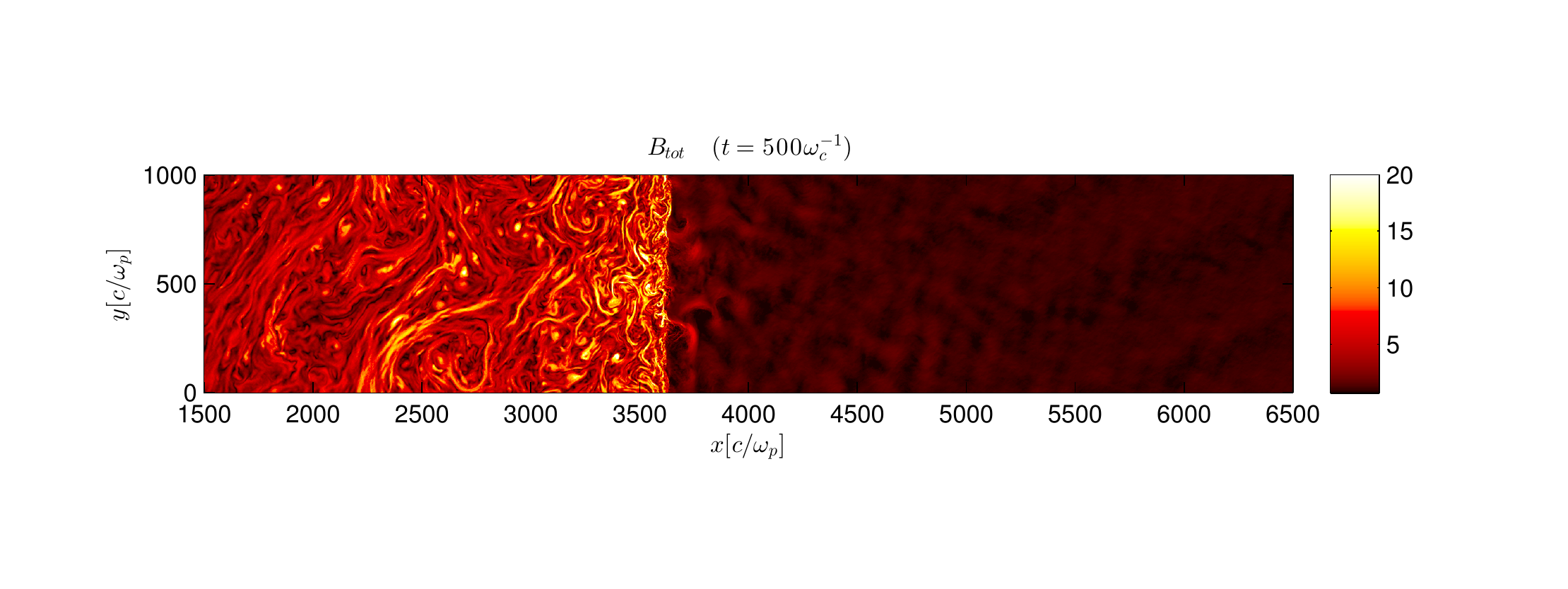}
\includegraphics[trim=40px 50px 40px 50px, clip=true, width=.80\textwidth]{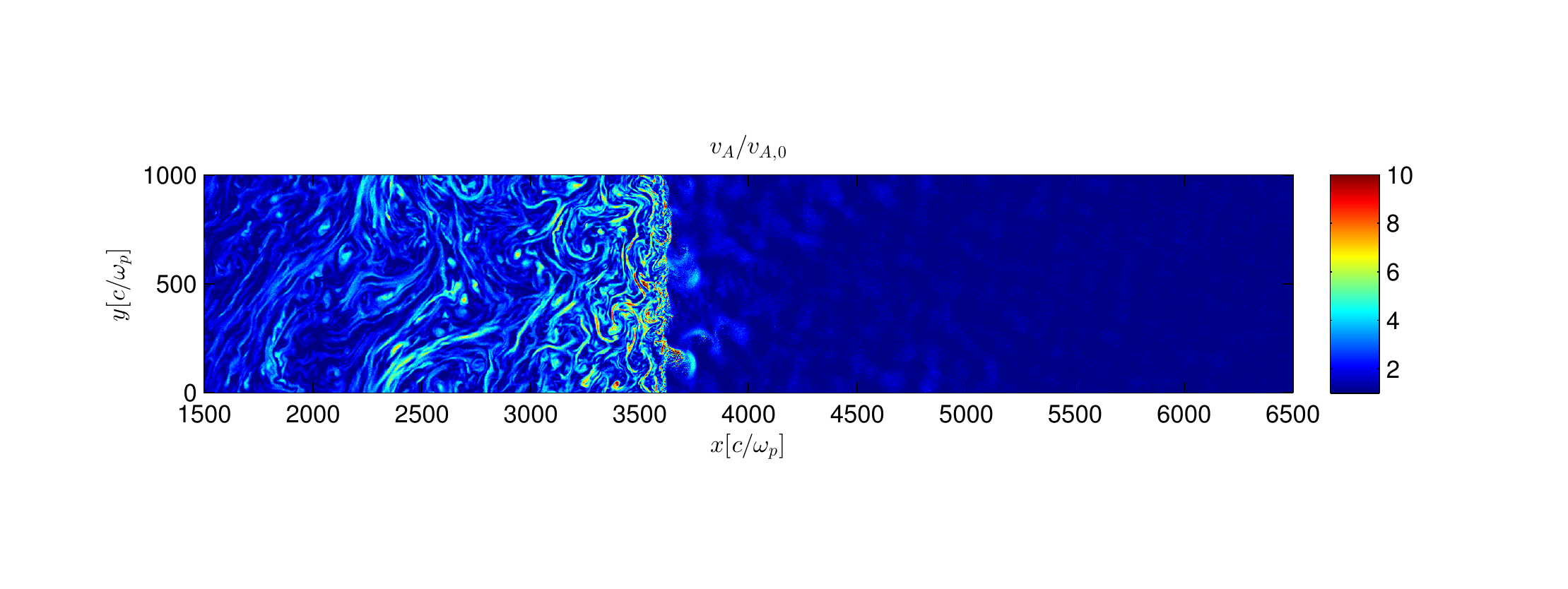}
\caption{\label{fig:th20}
As in Figure~\ref{fig:par}, but for an oblique shock with ${\bf B}_0$ in the $xy$-plane, at an angle $\vartheta=20^{\circ}$ with respect to ${\bf v}_{sh}$.}
\end{figure*}

Most of the simulations presented here are for a 2D parallel shock (i.e., background field ${\bf B}_0$ aligned with the shock velocity ${\bf v}_{sh}=v_{sh}\hat{x}$) with Alfv\'enic Mach number $M_A=v_{sh}/v_A=30$ in the downstream (simulation) reference frame.
Ions are initialized with temperature $T=mv_A^2/k_B$, in thermal equilibrium with electrons.  
In the 2D simulations, we include both in-plane and out-of-plane components of the ion momenta and  electromagnetic fields.

In our non-relativistic framework, velocities are normalized to the Alfv\'en speed $v_A$,  lengths to the ion skin depth $c/\omega_{p}=v_A/\omega_c$, and times to $\omega_c^{-1}$, with $\omega_{p}=\sqrt{4\pi n_0e^2/m}$ and $\omega_c=eB_0/mc$ being the ion plasma and cyclotron frequencies ($m$ is the proton mass). 
Density and magnetic fields are measured in units of the upstream initial values, $n_0$ and $B_0$.
The ion skin depth is resolved with two grid cells and the computational box measures $(L_x\times L_y)=10^4\times10^3 (c/\omega_p)^2$. 
The time step is chosen as $\Delta t =0.001\omega_c^{-1}$ for keeping the Courant number small and enhancing energy conservation.
The shock is generated by introducing a perfectly reflecting wall (left in the figures) that initially produces counter-streaming particles.  
This configuration is unstable and the system promptly forms a propagating sharp discontinuity with width of a few gyroradii of the downstream thermal particles, behind which the cold upstream flow is isotropized. 

The global evolution of the shock obtained with the present setup is showed in Figure~\ref{fig:rho}, where the density is plotted at different times.
The shock propagates to the right along $x$ axis.
Since $M\gg1$, the ratio between the plasma density behind and ahead of the shock rapidly approaches 4, and the post-shock fluid comes to rest in the wall (downstream) frame, dissipating the inflowing kinetic energy into heat.

Figure~\ref{fig:spectra} shows a snapshot of the ion distribution function at $t=500\omega_c^{-1}$, and the time evolution of the downstream spectrum.
According to first-order Fermi mechanism, particles scattered back and forth across the shock get accelerated effectively: the ion energy spectrum quickly develops a non-thermal power-law tail that, for the chosen parameters, comprises $\sim 15\%$ of the total ion energy. 
The details of the ion energy spectrum will be discussed in a forthcoming paper.

The peak of the Maxwellian distribution is consistently shifted to lower temperatures by about the same amount with respect to the case without accelerated particles.
Since all the particles are coupled through electromagnetic interactions, the pressure in non-thermal ions propagating upstream slows down the incoming fluid producing a precursor (top panel in Figure~\ref{fig:spectra}), a distinctive feature of shocks modified by the back-reaction of accelerated particles \citep[see, e.g.,][]{malkov-drury01}.

\subsection{Filamentation instability}
The most noticeable feature in Figure~\ref{fig:rho} is the formation of low-density  cavities ahead of the shock ($n/n_0\lesssim 10^{-2}$), whose origin can be explained in the following way.
Particles accelerated at the shock propagate upstream against the incoming fluid (top panel of Figure~\ref{fig:spectra}).
At first, the magnetic field is the background ${\bf B}_0$ only; therefore, ions do not diffuse in pitch angle, giving rise to a current ${\bf J}\parallel {\bf B_0}$. 
However, the super-Alfv\'enic streaming of the accelerated particles tends to excite transverse magnetic modes via plasma instabilities, eventually seeding the upstream medium with a transverse ${\bf \delta B}$.
The resulting Lorentz force ${\bf F}_L\propto-{\bf J}\times \delta {\bf B}$ pushes the plasma (and the frozen-in field) away from the region where the current is stronger, also focusing the energetic particles and helping to sustain the instability \citep{bell05,rb12}.
The net result is the formation of low-density tunnels filled with supra-thermal ions, modulated by the period of the underlying magnetic perturbation (see, e.g., the snapshot at $t=300\omega_c^{-1}$ in Figure~\ref{fig:rho}).

A sizable fraction of the energy in the magnetic turbulence is stored on scales $\gtrsim 100 c/\omega_p$, i.e., larger than the gyroradius $\sim v_{sh}/\omega_c=30 c/\omega_p$ of ions with velocity $\sim v_{sh}$.
Fastest-growing modes are predicted to be short-wavelength, non-resonant \citep{bell04,bell05,ab09}; 
therefore, magnetic field structures with length scales comparable to the gyroradii of the accelerated particles can be present only if Bell's modes grow to larger scales within an advection time, or if other classes of long-wavelength modes are excited as well.
We will discuss the properties of the CR-induced turbulence in greater detail in a forthcoming work.

\subsection{Growth of cavities and filaments}
An interesting question is what determines the size of the cavities.
\cite{rb12} argued that the growth of the cavities is suppressed when their size becomes comparable with the gyroradius of the ions carrying most of the current, since the scattering of these particles would suppress the instability. 
In our setup the ion current is both time- and space-dependent, and filamentation may also be limited by advection. 
While advected with the fluid, cavities probe the upstream profile of the ion current; also, the advection time may provide a limitation to the maximum size achievable by expanding cavities.

In our simulations cavities grow while being advected toward the shock and, by comparing different panels of Figure~\ref{fig:rho}, we notice that the typical transverse size of the cavities impacting the shock increases with time.
This is qualitatively consistent with a cavity growth rate $\Gamma\propto |B_{\perp}|\xi_{acc}^{1/2}$, where $|B_{\perp}|$ is the averaged magnitude of transverse component of ${\bf B}$, and $\xi_{acc}$ is the fraction of the total pressure in accelerated particles \citep{rb12}. 
In our simulations, $\xi_{acc}$ and, in turn, $|B_{\perp}|$ increase with time (see the non-thermal tail in the spectra of Figure~\ref{fig:spectra}).
At $t\lesssim 400\omega_c^{-1}$ the transverse size of the biggest cavity measures $\approx 300c/\omega_p$, and it is comparable with the gyroradius of the ion with the highest energy, $E_{max}\sim 100 E_{sh}$, with $E_{sh}=mv_{sh}^2/2$ (Figure~\ref{fig:spectra}).  

Lower-energy particles resonate with the self-generated magnetic turbulence and are effectively scattered, as shown in the top panel of Figure~\ref{fig:spectra}.
The density of ions with $E\lesssim 10E_{sh}$ drops with the distance from the shock: if these particles were not deflected back, they would escape the box on a timescale shorter than the simulation one.
By undergoing multiple interactions with the shock, ions are efficiently accelerated, and $E_{max}(t)$ increases with time.
On the other hand, freshly accelerated ions with energies $\sim E_{max}$ do not have enough waves to resonate with and stream more freely, contributing the most to the current because of their anisotropy.

For $t\gtrsim 500\omega_c^{-1}$, $E_{max}(t)$ no longer increases due to the finite size of the box in the $x$ direction: however, we checked that this limitation is removed for larger boxes and consistently longer simulation times.
Enlarging the box in the $y$ direction, instead, does not accommodate larger and larger cavities, attesting to the physical origin of their diameter.
The transverse size of the box must, however, be large enough to resolve the gyration of the most energetic ions at any time: the smaller the box in the $y$ direction, the earlier filamentation is suppressed.
Previous simulations \citep[e.g.,][]{Giacalone04,gs12} used smaller boxes, and, therefore, the effect was less evident.
\begin{figure*}\centering
\includegraphics[width=.80\textwidth]{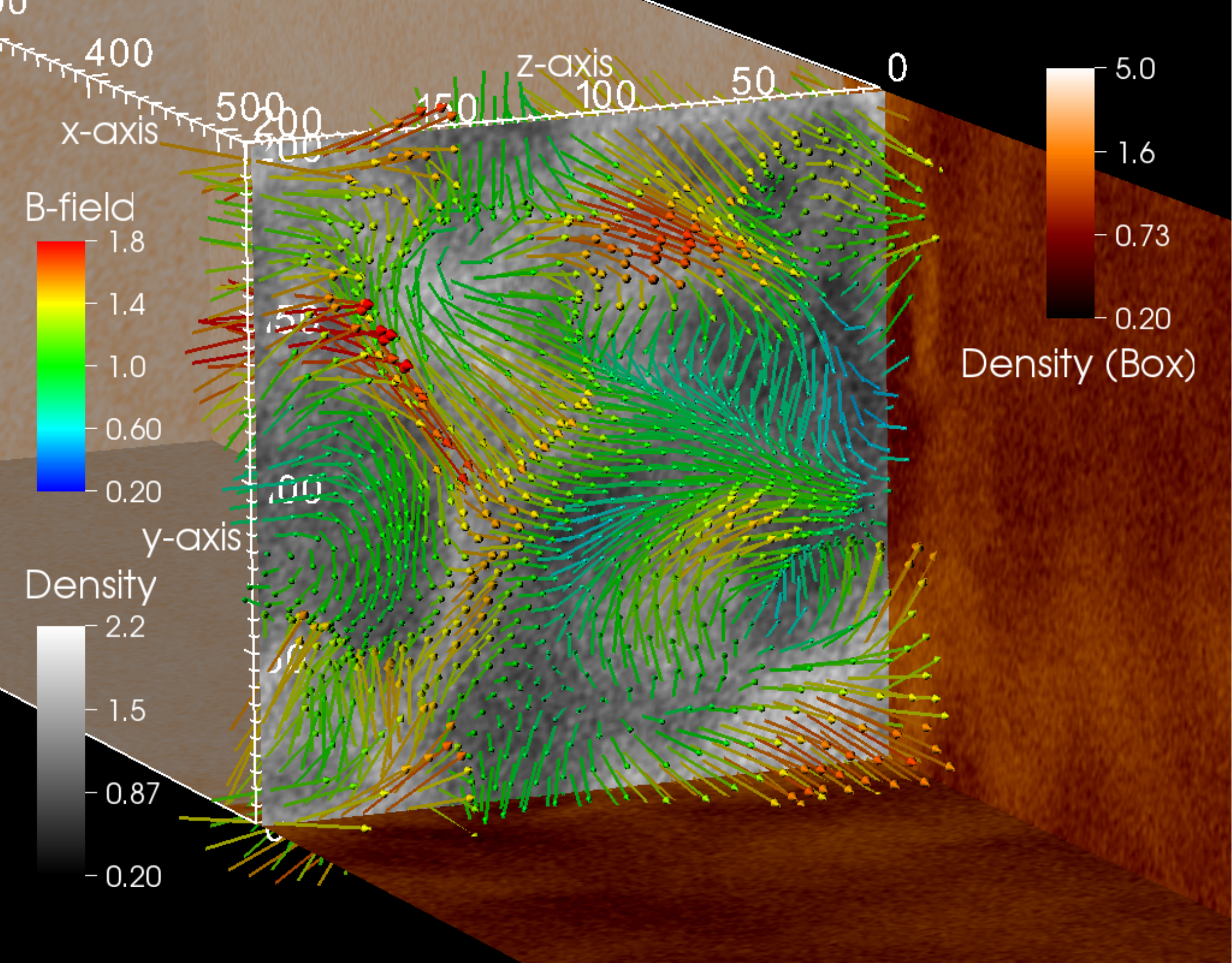}
\caption{\label{fig:3D} 
Snapshot at $t=175\omega_c^{-1}$ of a 3D hybrid simulation of a parallel shock with $M_A=6$ in a $2000\times 200\times 200 (c/\omega_p)^3$ box (see text for further details). 
The color code (right colorbar) on the box sides shows the particle density in units of $n_0$, while the vertical slice illustrates a section of the fluid at $x=520c/\omega_p$, i.e., immediately ahead of the shock.
In the slice, the grey-scale code corresponds to the ion density, while the colored vectors show strength and direction of the magnetic field, in units of $B_0$.
Notice the correlation between underdensity and low $B$-field and how the magnetic field is mainly along $\hat{x}$ in the filaments and coiled inside the cavities.
\emph{An animation is available in the online journal}.
}
\end{figure*}

\subsection{Magnetic field structure}
In Figure~\ref{fig:par} the parallel ($B_x$), transverse ($B_y$,$B_z$) and total ($B_{tot}$) magnetic fields are shown for the 2D case at $t=500\omega_c^{-1}$.
The filamentary structure of the upstream is very evident in $B_x$, where organized lateral modulations significantly affect the mean magnetic field topology.
In the non-linear stage of the instability, $B_x$ tends to be evacuated from cavities and pushed into filaments \citep[see][]{rb12}, while the transverse component winds up around the cavity itself.
Also, accelerated ions are focused into the cavities, contributing to the current that drives the instability. 
For instance, the current $J_x$ inside the most prominent upstream cavity (approximately for $400 c/\omega_p<y<550 c/\omega_p$ and $x>3800c/\omega_p$ in Fig.~\ref{fig:par}) is about a 50\% larger than in the densest filament (at $550 c/\omega_p<y<700 c/\omega_p$). 

There is also another interesting topological effect that arises in the scenario presented here. 
When a cavity is advected through the shock, a bubble of relatively rarefied and cold (it contains less kinetic energy to be dissipated) plasma is produced in the downstream (see, e.g., Figure~\ref{fig:rho} at $t=300\omega_c^{-1}$). 
Such a configuration is prone to the Richtmeyer--Meshkov instability, and each bubble is promptly filled by a \emph{plume} of hotter and denser plasma, as visible in Figure~\ref{fig:rho} between $t=400$ and 500$\omega_c^{-1}$.
These macroscopic motions systematically stretch the magnetic field along the cavity axis, as can be seen in $B_x$ and $B_{tot}$ in Figure~\ref{fig:par}.

A similar phenomenology is recovered for oblique shocks, too.
Figure~\ref{fig:th20} shows the output of a simulation for a shock with an angle of $20^\circ$ between ${\bf B}_0$ and ${\bf v}_{sh}$, and all the other parameters fixed as in the parallel case.
We note that the repeated cavity-filament pattern always develops along the direction of the upstream background field, being driven by energetic particles streaming along ${\bf B}_0$.
However, in this case the shock is less corrugated, and downstream cavities are squeezed more rapidly, preventing the formation of very long filaments.

\subsection{3D simulations}
To better understand the topology of these structures, we also ran 3D simulations with {\it dHybrid}.
Both the shock velocity and the computational domain are scaled down by a factor of 5 with respect to the 2D case to compensate for the higher computational effort without changing the ratio of the box size and the gyroradius of typical particles, $r_L\propto v_{sh}$. 
More precisely, we set $M_A=6$ in a 2000$\times 200\times 200 (c/\omega_p)^3$ box, allowing for the same time and space resolution as in the 2D case. 

Figure~\ref{fig:3D} shows the rendering of the density and field structure near the shock in a 3D simulation. 
The slice shows a transverse section of the fluid immediately ahead of the shock, where the contrast between filaments and cavities may be as large as 5--10 in both the ion density (grayscale) and in total magnetic field (color-coded vectors).
There is a clear correlation between regions of enhanced magnetic field and enhanced density, confirming the formation of rarefied, low-$B$ cavities surrounded by regions of denser plasma permeated by a stronger field.

An interesting feature clearly visible in 3D simulations is that the field is mainly parallel to the shock normal in the dense filaments, while inside the cavities the field is coiled and develops a transverse component comparable with, and even larger than the one along $\hat{x}$ \citep[see also][]{rb12}.

\section{Observational consequences}\label{sec:obs}
Very generally, we can assess that filamentation provides substantial magnetic field amplification even in the pre-shock medium.
The magnetic-field strength in some filaments may become as large as $\sim5 B_0$, while cavities may become effectively demagnetized.
While low-energy accelerated ions are focused in the inner regions of the cavities, ions with gyroradii comparable with the cavity transverse size can feel the effect of the field amplification and scatter more effectively.
The bottom panels of Figures~\ref{fig:par} and \ref{fig:th20} show that the local Alfv\'en velocity becomes significantly larger than the initial one, as a consequence of the increase of $B$ and/or the decrease of $n$.
The phase velocity of the magnetic perturbations may become a non-negligible fraction of the fluid velocity even for strong shocks, eventually affecting the scattering and, in turn, the spectra of the accelerated particles.
In particular, the effect may be crucial to explain the steep inferred from $\gamma$-ray observations of SNRs \citep{efficiency}. 

Also, the thermal plasma is affected by the filamentation process: ahead of the shock the temperature becomes several times larger than at the beginning of the simulation.
Quite interestingly, the pressure in the thermal plasma and in the magnetic turbulence are almost in equipartition along the precursor.

Due to the non-linear processes developing in the upstream, the magnetic field orientation and the local Alfv\'enic and sonic Mach numbers may vary significantly immediately ahead of the shock (Figures \ref{fig:par} and \ref{fig:th20}); 
different patches of the upstream plasma are therefore shocked in different ways, eventually leading to the onset of coherent motions that stretch the field preferentially along the cavities, but also turbulent motions that effectively stir the downstream fluid on smaller scales.

The net result is that the post-shock magnetic field may become even larger than what is expected from a simple compression by a factor $\sim 4$ of the transverse component of the pre-shock field.
In Figure~\ref{fig:par}, in addition to an elongated filament where $B\ge 15-20B_0$, it is possible to spot knots with $B\sim 30-40B_0$.
Runs with larger Mach numbers ($M_A=50$) show that the total $B$-field in knot-like structures can easily reach up to 90--100$B_0$. 
These structures may resemble the ones detected in RX J1713.7-3946, where the fields inferred by the X-ray variability are as strong as 1mG \citep{uchiyama+07}, namely a few hundred times stronger than the typical interstellar magnetic field.

In application to SNRs, if the cavity sizes continue to grow to the gyroradius of ions with $E_{max}$, filamentation instability might also account for the detection of a characteristic pattern of X-ray-bright stripes \citep{eriksen+11} in the South-East region of Tycho \citep[see also][]{bykov+11}.
In this case, stripes would develop where the shock normal is parallel to the large-scale magnetic field, and the stripe spacing would correspond to the gyroradius of protons with energy $\sim10^6$GeV, a value compatible with the maximum energy achieved in Tycho as inferred from $\gamma$-ray data \citep{Tycho}.

Though the hybrid simulations presented here cannot properly reproduce the large physical scales relevant for real SNRs, they qualitatively confirm that the instabilities driven by the accelerated particles can lead to the formation of filaments and knots sharing some properties --- such as the geometry and the enhancement of the magnetic field --- consistent with several observational signatures detected in young SNRs.

\subsection*{}
We wish to thank L. Gargat\'e for providing a version of \emph{dHybrid} and for his kind help, M. Kunz for having read a preliminary version of the paper, and the anonymous referee for her/his accurate comments. 
This research was supported by NSF grant AST-0807381 and NASA grants NNX09AT95G and NNX10A039G. Simulations were performed on the computational resources supported by the PICSciE-OIT TIGRESS High Performance Computing Center and Visualization Laboratory. This research also used the resources of the National Energy Research Scientific Computing Center, which is supported by the Office of Science of the U.S. Department of Energy under Contract No. DE-AC02-05CH11231, and Teragrid/XSEDE's Ranger under contract No.\ TG-AST100035.

\bibliographystyle{yahapj}
\bibliography{filaments}

\end{document}